\documentclass[10pt,aps,prd,twocolumn,twoside,superscriptaddress,floatfix, nofootinbib,amssymb]{revtex4-1}
\usepackage[T1]{fontenc}

\pdfoutput=1
\usepackage{ifthen,amsmath,amssymb, bm}
\usepackage{graphicx}
\usepackage{xcolor}
\usepackage{hyperref}
\usepackage{float}
\usepackage{aas_macros}
\usepackage[compatibility=false]{caption}
\usepackage{subcaption}

\newcommand{\beq} {\begin{equation}}
\newcommand{\eeq} {\end{equation}}
\newcommand{\bal} {\begin{aligned}}
\newcommand{\eal} {\end{aligned}}

\begin{document}

\title{
Exploring gas thermodynamics around galaxies from the Sunyaev-Zel'dovich effects:
impact of galaxy-halo connection, 2D projection and velocity field\\
}

\author{Sadaf Kadir}
\email{safkadir@stanford.edu}
\affiliation{Department of Physics, Stanford University, Stanford, CA, USA 94305-4085}

\author{Bernardita Ried Guachalla}
\email{bried@stanford.edu}
\affiliation{Department of Physics, Stanford University, Stanford, CA, USA 94305-4085}
\affiliation{Kavli Institute for Particle Astrophysics and Cosmology, 382 Via Pueblo Mall Stanford, CA 94305-4060, USA}
\affiliation{SLAC National Accelerator Laboratory 2575 Sand Hill Road Menlo Park, California 94025, USA}

\author{Sihan Yuan}
\affiliation{Kavli Institute for Particle Astrophysics and Cosmology, 382 Via Pueblo Mall Stanford, CA 94305-4060, USA}

\author{Emmanuel Schaan}
\affiliation{Kavli Institute for Particle Astrophysics and Cosmology, 382 Via Pueblo Mall Stanford, CA 94305-4060, USA}
\affiliation{SLAC National Accelerator Laboratory 2575 Sand Hill Road Menlo Park, California 94025, USA}

\author{Risa H. Wechsler}
\affiliation{Department of Physics, Stanford University, Stanford, CA, USA 94305-4085}
\affiliation{Kavli Institute for Particle Astrophysics and Cosmology, 382 Via Pueblo Mall Stanford, CA 94305-4060, USA}
\affiliation{SLAC National Accelerator Laboratory 2575 Sand Hill Road Menlo Park, California 94025, USA}

\begin{abstract}

A complete picture of the gas thermodynamics around galaxies is imprinted on the cosmic microwave background (CMB).
Indeed, the thermal, kinematic, and relativistic Sunyaev-Zel'dovich  effects (tSZ, kSZ, rSZ), along with screening and CMB lensing, measure the gas density, temperature, pressure of baryonic feedback and bulk velocity around galaxies, along with the gravitational potential it sits in.
This full thermodynamic picture promises to constrain galaxy formation models and gas related uncertainties in the impact on galaxy lensing.
Recent kSZ measurements around galaxies suggest that the gas may be more extended than anticipated, pointing to powerful feedback processes and large baryonic corrections to lensing.
How robust are these conclusions about the galaxy–halo connection, including satellite fraction and high-mass outliers, or to 2D projection effects and large-scale velocity modes?
In this paper, we give simple estimates for these effects using a simulated sample of DESI-like luminous red galaxies within the IllustrisTNG hydrodynamical simulation and the Abacus $N$-body simulation.
We show that analyzing projected 2D profiles, rather than the intrinsic 3D gas and matter distributions, can lead to biases when computing quantities like the gas fraction.
We also find that in the absence of spatial filtering, the 2-halo term is non-negligible for kSZ even at the smaller radii where the 1-halo term dominates.
We show that a 1\% uncertainty in the satellite fraction of galaxies can propagate into uncertainties of $\pm 1\%, \pm 3\%$ and $\pm 5\%$ in the 1-halo terms of the kSZ, tSZ, and rSZ signals, respectively.
We show that masking the 2\% most massive objects in the sample reduces the profile amplitudes by up to 10\%, 40\%, and 75\% for the kSZ, tSZ, and rSZ signals, respectively.
Finally, we show that na\"ive simulations of the kSZ effect can be biased by an artificial Doppler term, which is automatically removed when high-pass or compensated aperture filtering is applied.

\end{abstract}

\maketitle


\section{Introduction}

Galaxies exhibit a complex interplay between star formation and quenching across cosmic time \cite{Somerville2015}, with feedback processes playing a central role in regulating this evolution \cite{Fabian2012}. 
Observationally, constraining the thermodynamic properties of gas in galaxies and groups has remained challenging, while hydrodynamical simulations often lack the resolution to capture the relevant small-scale physics, relying instead on sub-grid models \cite{Angelinelli2023, vanDaalen2011}. 
With the advent of high-precision cosmological surveys, such as the Rubin Observatory \cite{lsstsciencecollaboration2009lsst}, DESI \cite{Aghamousa2016}, Euclid \cite{Amendola_2018}, and the Nancy Grace Roman Space Telescope \cite{Spergel_2015}, an incomplete understanding of baryonic physics can directly affect the interpretation of clustering and lensing measurements \cite{Semboloni_2011}.

The Sunyaev–Zel’dovich (SZ) effects \cite{Zeldovich_Sunyaev_1969}, arising from the inverse Compton scattering of cosmic microwave background (CMB) photons by free electrons, provide a powerful tool to probe the thermodynamics of the circumgalactic medium. 
Thermally energized electrons transfer energy to CMB photons, producing a frequency-dependent distortion known as the thermal SZ (tSZ) effect \cite{Sunyaev_Zeldovich_1972}, which directly traces the gas thermal pressure. 
Alternatively, the bulk motion of a halo along the line of sight induces a Doppler shift in scattered CMB photons, giving rise to the kinematic SZ (kSZ) effect \cite{Sunyaev1980}, which is proportional to the gas momentum along the line of sight. 
Together, the tSZ and kSZ effects provide complementary information about the thermal and dynamical state of gas in galaxies and their halos \cite{Battaglia2019Probing}. 
Thanks to advances in CMB instrumentation, telescopes such as ACT \cite{Fowler_2007, Swetz2011, Thornton_2016, Naess_2020}, SPT \cite{Carlstrom2011}, the Simons Observatory \cite{Ade2019}, and upcoming CMB-S4 experiments \cite{Abazajian2016} will dramatically improve the precision of SZ measurements in the coming decade \cite{Battaglia_2017}.

Recent stacking analyses of the tSZ and kSZ around DESI luminous red galaxies (LRGs) have provided unprecedented constraints on feedback processes, revealing that baryonic gas extends to megaparsec scales, well beyond the central regions of dark matter halos \cite{hadzhiyska2024evidencelargebaryonicfeedback, riedguachalla2025backlightingextendedgashalos}. 
Together with earlier measurements \cite{Schaan:2015uaa, Schaan2021}, these studies are increasingly being used to calibrate the impact of baryons on weak-lensing observables and matter power spectrum suppression in both analytic models and hydrodynamical simulations \cite{Amodeo_2021, Schneider2022, bigwood2024weak, mccarthy2024flamingocombiningkineticsz, sunseri2025disentanglinghalojointmodel, bigwood2025kineticsunyaevzeldovicheffect, siegel2025jointxraykineticsunyaevzeldovich, Wayland2025, wayland2025detailedtheoreticalmodellingkinetic}. 
Joint tSZ–kSZ analyses also provide insight into halo thermodynamics and related processes, including star formation, infrared emission, and active galactic nucleus activity \cite{Battaglia_2017, Meinke2021, Tanimura_2021, Vavagiakis_2021}. 
In the future, patchy screening of CMB photons will further enhance our understanding of thermodynamics of the gas, as shown in \cite{Schutt2024, coulton2025atacamacosmologytelescopesearch, Hadzhiyska2025_ksz_patching}

In this work, we aim not to produce a definitive model of gas profiles, but rather to use a plausible hydrodynamical simulation, IllustrisTNG, to gain insight into key systematic effects that impact SZ measurements. 
Building on previous simulation studies \cite{Springel2001, Refregier2002,Dolag2016,Moser2021,Moser2022,Moser2023,Hadzhiyska2023,Hill_2018,popik2025impactshalomodelimplementations}, we model the tSZ (including its relativistic correction) and kSZ profiles for DESI-like LRGs, deconstructing the 1-halo and 2-halo contributions, and examining the roles of centrals and satellites. 
We further investigate the impact of projecting 3D gas distributions into 2D profiles, the influence of large-scale velocity modes on kSZ, and whether the IllustrisTNG box size is sufficient to model observed 2D profiles. 
We also assess the extent to which SZ stacking techniques effectively remove the Doppler component of the kSZ effect.

Through this analysis, we aim to quantify the magnitude of these systematic effects and inform future modeling strategies for accurately recovering the thermal properties of galaxies. 
This study serves as a preparatory exercise to identify the most critical factors that must be addressed to fully exploit the higher-precision SZ measurements expected from upcoming surveys. 
In Section~\ref{sec:theory}, we describe the theoretical background of the SZ effect; 
Section~\ref{sec:methods} outlines our methodology, including the SZ stacking procedure and the simulated DESI-like sample; 
Section~\ref{sec:results} presents the resulting profiles, and Section~\ref{sec:conclusions} summarizes our conclusions.

\section{Review of the Sunyaev-Zel'dovich effects}
\label{sec:theory}

The SZ effect is a spectral distortion in the CMB resulting from inverse Compton scattering of CMB photons by electrons in the circumgalactic and intracluster medium, where the electrons are hot and ionized.
In the non-relativistic case (lowest order in electron velocities), the photons undergo a frequency-dependent temperature shift, denoted as $\Delta T_{\rm tSZ}$, given by \cite{Rephaeli1995, Carlstrom2002}:
\begin{equation}
    \frac{\Delta T_{\rm tSZ}}{T_\text{CMB}} =f_{\rm 0}(x)\ y_{\rm tSZ},
\end{equation}
where the dimensionless frequency is defined as
\begin{equation}
    x \equiv \frac{h\nu}{k_{\rm B} T_\text{CMB}},
\end{equation}
with $h$ the Planck constant, $\nu$ the frequency, $k_{\rm B}$ the Boltzmann constant, and $T_\text{CMB}$ the CMB temperature.
The frequency dependence of the tSZ can be derived from first principles \cite{Carlstrom2002} as:
\begin{equation}
    f_{\rm 0}(x) =  x \left(\frac{e^x +1}{e^x-1}\right) - 4 .
\end{equation}
The Compton $y$-parameter $y_{\rm tSZ}$, which determines the amplitude of the effect, is defined as:
\begin{equation}
    y_{\rm tSZ} 
    \equiv 
    \sigma_{\rm T} \int n_{\rm e}(\ell) \frac{k_{\rm B} T_{\rm e}(\ell)}{m_{\rm e} c^2} d\ell
    \label{eq:compton_tSZ}
\end{equation}
with $\sigma_{\rm T}$ the Thomson cross-section, $n_{\rm e}$ the electron number density, $T_{\rm e}$ the electron temperature, $m_{\rm e}$ the rest mass of the electron, $c$ the speed of light and the integral goes along the line-of-sight (LOS) physical distance $d\ell$. 
In particular, the Compton parameter measures the LOS integral of the electron thermal pressure:
$y_{\rm tSZ} \propto n_{\rm e} T_{\rm e} \propto P_{\rm e}$.
Therefore, the resulting integral of the Compton $y$ parameter over the 2D angular area of a cluster scales as $\int y_{\rm tSZ} \hspace{0.1 cm} d^2\theta \propto M_{\rm h}^{5/3}$, with $M_{\rm h}$ being the mass of the halo.

At the next order in electron velocities, a relativistic correction arises. 
The corresponding additional temperature shift, denoted as $\Delta T_{\rm rSZ}$, is given by:
\begin{equation}
    \frac{\Delta T_{\rm rSZ}}{T_\text{CMB}} =f_{\rm 1}(x)\  y_{\rm rSZ}
\end{equation}
with
\begin{equation}
    y_{\rm rSZ} = \sigma_{\rm T} \int n_{\rm e}(\ell) \left(\frac{k_{\rm B} T_{\rm e}}{m_{\rm e} c^2}\right)^2 d\ell,
    \label{eq:compton_rSZ}
\end{equation}
and corresponding frequency dependence is:
\begin{align*}
    f_{\rm 1}(x) = 
    &\frac{7}{10} x^3 \left(\frac{e^x +1}{e^x-1}\right)^3 - \frac{42}{5} x^2 \left(\frac{e^x +1}{e^x-1}\right)^2 \\
    & + \frac{47}{2} x \left(\frac{e^x +1}{e^x-1}\right) - 10 \\
    & + \frac{x^2}{\sinh^2{(x/2)}} \left(\frac{7}{5} x \left(\frac{e^x +1}{e^x-1}\right) - \frac{21}{5} \right),
\end{align*}
as derived in Eq.~2.27 of \cite{Itoh1998}.
The relativistic correction thus scales with an additional power of temperature, $y_{\rm rSZ} \propto n_{\rm e} T_{\rm e}^2$.
Therefore, the resulting integral of $y_{\rm rSZ}$ over the cluster area scales as $\int y_{\rm rSZ} \hspace{0.1 cm} d^2 \theta \propto M_{\rm h}^{7/3}$.

Alternatively, when the free electrons have a non-zero bulk motion with respect to the CMB, the Doppler shift of CMB photons introduces an additional temperature shift, which we denote as $\Delta T_{\rm kSZ}$. 
This shift is given by:
\begin{align}
    \frac{\Delta T_{\rm kSZ}}{T_{\rm CMB}} 
    & = - \sigma_{\rm T} \int n_{\rm e}(\ell) e^{- \tau} \frac{v_{\parallel}}{c} d\ell, \\
    & \approx - \tau \frac{v_{\parallel}}{c}, 
    \label{eq:kSZ}
\end{align}
where $\tau = \int n_{\rm e} \sigma_{\rm T} d\ell$ is the dimensionless Thomson optical depth,
and $v_{\parallel}$ is the LOS peculiar bulk velocity of the free electron gas.
We have used that $\tau \ll 1$, meaning that $e^{- \tau} \approx 1$.
For one cluster in particular, the integral of $\tau$ over the cluster area scales as $\int \tau \hspace{0.1 cm} d^2 \theta \propto M_{\rm h}$.
Throughout this work, we will study the SZ relevant quantities: $y_{\rm tSZ}$, $y_{\rm rSZ}$ and $\tau$, respectively.

The kSZ effect can be split into two terms.
We decompose the electron number density into a mean number and an overdensity
\begin{equation}
    n_e = \overline{n}_e (1+ \delta),
\end{equation}
which allows us to split the kSZ effect into two components:
\begin{align}    
    \frac{\Delta T_{\rm kSZ}}{T_{\rm CMB}} &= - \int  \overline{n}_e  \ \frac{v_{\parallel}}{c} \sigma_{\rm T} \ d\ell - \int \overline{n}_e \delta \ \frac{v_{\parallel}}{c} \sigma_{\rm T} \ d\ell  \\
    & = \frac{\Delta T_{\rm Doppler}}{T_{\rm CMB}} + \frac{\Delta T_{\rm OV}}{T_{\rm CMB}}
    \label{eq:kSZ_doppler_other}
\end{align}
where the first component corresponds to the ``Doppler'' term ($\Delta T_{\rm Doppler}$) and the second one corresponds to the ``Ostriker-Vishniac'' term ($\Delta T_{\rm OV}$) \cite{Ostriker1986, Vishniac1987}.
The Doppler term  describes large-scale velocity anisotropies, while the Ostriker-Vishniac term accounts for gas density fluctuations weighted by their peculiar velocities \cite{Alvarez2016}.
We had assumed that the optical depth is small ($e^{-\tau} \approx 1$).

In the linear regime of the peculiar velocity field, the Doppler term, when integrated over large distances, is expected to be strongly subdominant relative to the Ostriker–Vishniac term. 
This is because the velocity can be expressed as the gradient of the matter density field, itself representable as a superposition of plane waves along the line of sight. 
In the absence of cosmic evolution, the contribution from one half of a wave cancels exactly with the other \cite{1978Sunyaev}. 
Evolution of the density and velocity fields breaks this symmetry, leaving a small residual contribution \cite{1984Kaiser, 2002Ma}.
A detailed historical review and derivation is provided in \cite{Alvarez2016}.
In a survey, the degree of cancellation of the Doppler term depends on its properties; 
for sufficiently long-wavelength modes, if the survey is not deep enough in redshift, it can leave a residual contribution to the large-scale velocity field, and consequently, to the kSZ signal. 
If the goal is to study the gas content of individual galaxies via the kSZ, it is therefore necessary to ensure that the Doppler contribution is negligible with respect to the Ostriker--Vishniac term.
We explore the extent of this scenario in more detail in Sec.~\ref{sec:doppler}.

When measuring the SZ profiles, we do not have direct access to the true center or the gas distribution, but only to the galaxies that reside within halos. 
As a result, galaxies act as biased tracers, introducing effects such as miscentering and potential double-counting when multiple galaxies belong to the same halo and this is not accounted for.
Therefore, the kSZ profiles derived from combining galaxies and CMB maps are effective overdensity profiles, i.e., the excess signal over the mean in the Universe along the LOS, traced by visible matter.
Importantly, the combined SZ effects provide insight into the complete thermodynamics of gaseous halos \cite{Battaglia:2016xbi, Amodeo2021}, as the tSZ, rSZ and kSZ together offer information about the integrated electron pressure, temperature and density of the gas.
However, the gas traced through the SZ profiles includes contributions from both the 1- and 2-halo terms, meaning the signals must be disentangled to accurately describe the thermodynamics of individual halos. 
Additionally, velocity effects, such as the Doppler term, can also leak into the kSZ signal, requiring extra care when interpreting the gas density inferred from kSZ.
The extent of these subtleties is further studied in this work.

\section{Methods}
\label{sec:methods}

\subsection{Galaxy-traced halo gas profiles: toward robust feedback constraints
}

While our analysis uses simulations, our goal is to gain insights into the systematic effects that impact SZ measurements of galaxy's gaseous halos. 
To this end, we mimic the stacking procedure applied in recent DESI Y1 analyses on ACT DR6 maps \cite{hadzhiyska2024evidencelargebaryonicfeedback, riedguachalla2025backlightingextendedgashalos, liu2025measurementsthermalsunyaevzeldovicheffect}, enabling us to explore how factors such as 2D projection, satellite fractions, halo mass selection, and simulation box size influence the recovered profiles.
This method allows us to probe the surrounding gaseous halos by using the galaxies they host as tracers.

The stacking SZ method consists of localizing galaxies in the 2D plane of the CMB map and stacked at their positions. 
For the tSZ signal, the stacked profile is obtained by simple averaging, whereas for the kSZ, each cutout is weighted by the peculiar LOS velocity of the galaxy. This procedure effectively cancels CMB foreground contamination, as galaxy motions toward and away from the observer are symmetric.
The DESI LRG stacking analyses employed the Compensated-Aperture Photometry (CAP) filter, which subtracts the SZ signal in an outer ring from that in an inner aperture:
\begin{equation}
  \mathcal{T}(R) =
\int_0^{R} d^2\theta \hspace{0.05 cm} \delta T(\theta)
- \int_{R}^{\sqrt{2} R} d^2\theta \hspace{0.05 cm} \delta T(\theta).
    \label{eq:CAP}
\end{equation}
By subtracting up to $\gtrsim$3 arcmin scales, the primary CMB contribution is removed.
The resulting tSZ stack corresponds to
\begin{equation}
    \hat{T}_{\rm tSZ} (R) = \frac{\sum_i^N \mathcal{T}_i(R)}{N}.
\label{eq:stack_tsz}
\end{equation}

For the kSZ, the peculiar velocities are estimated from the DESI Y1 Baryonic Acoustic Oscillations reconstruction \cite{DESI2024.III.KP4, 2024Paillas}, which uses the galaxy number density field and calculates the displacement field inferred from the linear continuity equation. 
From this, it is straightforward to obtain a velocity field by taking the gradient of the displacements \cite{Ried_Guachalla_2024, hadzhiyska2023syntheticlightconecatalogues}.
The resulting kSZ stack is equivalent to
\begin{equation}
    \hat{T}_{\rm kSZ} (R)
    = - \frac{1}{r} 
    \frac{\sigma_v^{\rm rec}}{c}
    \frac{\sum_i^N \mathcal{T}_i(R) (v^{\rm rec}_i/c)}{\sum_i^N (v^{\rm rec}_i/c)^2},
\label{eq:stack_ksz}
\end{equation}
where $v^{\rm rec}_i$ are the reconstructed LOS velocities, $r$ is the correlation coefficient between true and reconstructed velocities, and $\sigma_v^{\rm rec}$ is the RMS of the reconstructed velocities \cite{hadzhiyska2023syntheticlightconecatalogues}. 
In simulations, the true velocities are given, so the reconstruction correction is not applied.

By implementing this stacking methodology in IllustrisTNG, we can study how observational and analysis choices propagate into the recovered SZ profiles. 
This allows us to quantify the impact of projection effects, the relative contributions of centrals and satellites, the influence of massive halos, and the role of simulation box size.

In Sec.~\ref{sec:sim_sample}, we describe our simulated DESI-like LRG sample, and in Sec.~\ref{sec:observables}, we present the resulting 2D SZ profiles and a series of consistency checks, illustrating which systematic effects are most important to consider when interpreting real measurements.


\subsubsection{Simulated DESI-like luminous red galaxies in \textsc{IllustrisTNG}}
\label{sec:sim_sample}

To investigate the impact of galaxy selection on SZ measurements, we employ a simulated catalog of DESI-like luminous red galaxies (LRGs) from the IllustrisTNG hydrodynamical simulation \cite{Yuan2022}, which provides a realistic representation of complex hydrodynamical processes.
It should be emphasized that, although we adopt a particular galaxy selection from a given simulation, we do not claim or assume that the derived SZ profiles presented in Sec.~\ref{sec:results} accurately represent the true LRG sample, nor do they reflect recent observed results from \cite{hadzhiyska2024evidencelargebaryonicfeedback, riedguachalla2025backlightingextendedgashalos, liu2025measurementsthermalsunyaevzeldovicheffect}. 
Instead, we use these profiles to explore the subtleties in the resulting SZ signals and to highlight potential complexities when comparing observations with simulations, and leave their comparison for future work.

For context, the corresponding properties of DESI LRGs \cite{Zhou_2023} are as follows: 
LRGs in the DESI survey exhibit low redshift errors and minimal stellar contamination \cite{Zhou_2023_2}.
These galaxies are predominantly centrals, with a satellite fraction around $11-14\%$ \cite{yuan2023desi}, resulting in reduced virial motions and smaller redshift-space distortions in the 1-halo term regime. 
LRGs form the primary target sample in the redshift range $0.4 < z < 1.1$, providing the strongest constraints in clustering analyses \cite{DESI2024.V.KP5}.

On the simulation side, IllustrisTNG \cite{Nelson_2017, Springel_2017, Marinacci_2018, Naiman_2018, nelson2021illustristngsimulationspublicdata} is a series of large cosmological magnetohydrodynamical simulations with a galaxy formation model built upon the previous Illustris simulation \cite{Nelson_2015, Pillepich2017}, incorporating e.g. black hole-driven feedback \cite{Weinberger2016}. 
The largest simulation box, IllustrisTNG-300, has a size of 205 cMpc/$h$ and contains 2,500$^3$ dark matter (DM) and gas cells, with masses of 3.98 $\times$ 10$^7$ $M_{\odot}/h$ for DM and 7.44 $\times$ 10$^6$ $M_{\odot}/h$ for baryonic mass particles. 
The IllustrisTNG-300-Dark counterpart, which evolves with the same initial conditions, uses a halo finder based on the friends-of-friends (FoF) algorithm (for central halos) and a subhalo finder based on the SUBFIND algorithm \cite{Springel2001} (for satellite halos). 
Smaller IllustrisTNG boxes are not considered, as their sizes (35 and 75 cMpc/$h$, respectively) are insufficient to sample the most massive LRG host halos, which contain at most a few dozen $10^{13}$ and $10^{14} M_{\odot}$ halos respectively at $z=0$ \cite{Nelson_2019, Pillepich_2017}.

The selection of LRG-like galaxies in IllustrisTNG, as described in \cite{Yuan2022}, aims to reproduce the DESI color-magnitude selection cut outlined in \cite{Zhou2020} at the effective redshift of the LRG sample ($z = 0.8$), using the galaxy color model from \cite{Hadzhiyska2021}. 
This method results in the selection of 4,608 DESI-like LRGs, along with their corresponding DM halos in IllustrisTNG-300-Dark.
The host halo masses are given as $M_{\rm 200c}$, the mass enclosed within a radius that has an overdensity equal to 200 times the critical density of the Universe.
The resulting simulated LRG catalog is validated by fitting a standard five-parameter Halo Occupation Distribution (HOD) model from \cite{Zheng2007}, with the clustering results from the DESI LRG target selection \cite{Zhou2020} showing good agreement.
The mean halo mass of the simulated sample corresponds to $\log_{10} \overline{M}_{\rm h} = 13.38$ in units of $M_{\odot}/h$ and the satellite fraction is $f_{\rm sat} = 0.20$.
These values differ somewhat from those inferred by fitting the same HOD model to the early DESI data release \cite{yuan2023desi}, which for the redshift range $0.6 < z < 0.8$ yields $\log_{10} \overline{M}_{\rm h} = 13.26 \pm 0.02$ in units of $M_{\odot}/h$ and $f_{\rm sat} = 10.4^{+0.013}_{-0.010}$.
While the halo masses are comparable, the elevated satellite fraction in the simulation is expected to amplify the influence of satellites in our analysis.

\subsubsection{
Measuring thermodynamical 2D and 3D profiles around simulated DESI luminous red galaxies
}
\label{sec:observables}

To compute the electron number density from the gas cells in the simulations, we assume a fully ionized environment with a fixed mass fraction of 76\% hydrogen and 24\% helium. 
The impact of metallicity and varying ionization fraction is ignored, as these effects are sub-percent in magnitude \cite{nelson2021illustristngsimulationspublicdata}.
The mean molecular weight of each gas cell corresponds to
\begin{equation}
    \mu = \frac{4}{1+3X_{\rm H}+4X_{\rm H}x_{\rm e}} m_{\rm p},
    \label{eq:mean_molecular_weight}
\end{equation}
where $X_{\rm H}=0.76$ is the hydrogen mass fraction, $x_{\rm e}= n_{\rm e}/n_{\rm H}$ is the number ratio of free electrons to hydrogen atoms, provided by IllustrisTNG, and $m_{\rm p}$ is the proton mass. 
We have also assumed 
$m_{\rm H} \approx m_{\rm p}$, $m_{\rm He} \approx 4 m_{\rm p}$ 
and 
$m_{\rm e} \ll m_{\rm p}$.
The gas temperature is inferred as
\begin{equation}
    T_{\rm e}=(\gamma-1)\frac{u}{k_{\rm B}}\mu,
    \label{eq:electron_temp}
\end{equation}
where $\gamma=5/3$ is the adiabatic index, and the specific internal energy per unit mass $u$ is provided by IllustrisTNG.

\subsection{Disentangling 1-halo \& 2-halo terms}

The SZ signal from halos can be decomposed into 1- and 2-halo terms, corresponding respectively to contributions from particles within the host halo and from neighboring halos projected along and across the line of sight. 
In practice, stacking at galaxy positions introduces miscentering compared to the halo center, yielding an averaged 1-halo profile that is slightly offset and also includes the 2-halo contribution. 
For simplicity, we ignore miscentering in this work, focusing on disentangling the SZ contributions from the 1- and 2-halo terms, with detailed modeling of miscentering deferred to future studies.

If we ignore redshift dependences, the cross-power spectra $P_{\rm h, SZ}$ between galaxies placed at the halos' center and the various SZ profiles as a function of halo mass $M_{\rm h}$ can be decomposed as:
\begin{equation}
P_{\rm g, SZ}(k) = P_{\rm g, SZ}^{1h}(k) + P_{\rm g, SZ}^{2h}(k).
\end{equation}
The 1-halo term is given by:
\begin{equation}
P_{\rm g, SZ}^{1h}(k) = \int dM_{\rm h} \hspace{0.1 cm} n(M_{\rm h}) \hspace{0.1 cm} \frac{\overline{N}_{\rm g}(M_{\rm h})}{\overline{n}_{\rm g}} \hspace{0.1 cm} {\rm SZ}(k, M_{\rm h}),
\label{eq:1-halo}
\end{equation}
where $n(M_{\rm h})$ is the halo mass function, $\overline{N}_{\rm g}$ is the mean number of galaxies at mass $M_{\rm h}$, $\overline{n}_{\rm g}$ is the mean galaxy number density, and ${\rm SZ}(k, M_{\rm h})$ denotes the Fourier transform of the SZ profile centered at the halo under consideration.

The 2-halo term accounts for the halo bias $b(M_{\rm h})$:
\begin{align}
P_{\rm g, SZ}^{2h}(k)
\approx & \hspace{0.1 cm} P^{\rm lin}(k) \int dM_{\rm h} \hspace{0.1 cm} n(M_{\rm h}) \hspace{0.1 cm} \frac{\overline{N}_{\rm g}(M_{\rm h})}{\overline{n}_{\rm g}} \hspace{0.1 cm} b(M_{\rm h}) \nonumber \\
& \times \int dM_{\rm h} \hspace{0.1 cm} n(M_{\rm h}) \hspace{0.1 cm} {\rm SZ}(k, M_{\rm h}) \hspace{0.1 cm} b(M_{\rm h}).
\label{eq:2-halo}
\end{align}

From Eq.~\ref{eq:1-halo}, we see that the mass dependence of the 1-halo term follows that of the SZ profile, described in Sec.~\ref{sec:theory}.
In contrast, in Eq.~\ref{eq:2-halo}, the 2-halo term is modulated by the halo bias, which is primarily determined by the large-scale environment of halos \cite{Shi_2017}, resulting in a shallower mass scaling.
For a more detailed analysis of the 2-halo term and the overall impact of the halo model assumptions in the cross-correlation with SZ profiles, we refer the reader to \cite{Hill_2018, popik2025impactshalomodelimplementations, harscouet2025kszeveryonepseudoclapproach, wayland2025detailedtheoreticalmodellingkinetic}.

In simulations, the stacked SZ profiles can be decomposed directly by assigning all particles within a specified radius around the halo’s center of mass to the 1-halo term, then subtracting this from the total profile to isolate the 2-halo contribution.
In the outskirts of halos, the assignment of particles to a halo can become somewhat arbitrary, and dependent on the halo finding algorithm.
As a result, the distinction between 1-halo and 2-halo terms below is most useful and well-defined when one of them really dominates over the other, rather than in the transition region where 1-halo and 2-halo terms are comparable.
In this work, we assign all gas cells within $R_{\rm 200c}$ of the matched host halo center to the 1-halo term while the 2-halo term is the residual.

\subsection{Simulation size required to capture the 2-halo term}

What size simulation is needed to fully capture the 1-halo and 2-halo terms of the SZ signals?
Even for small radial distances away from galaxies and halos, the projected SZ signals may receive contributions from large distance along the LOS.
In this subsection, we give simple, heuristic estimates for the typical simulation box size needed for the various profiles considered: total mass, gas mass, tSZ, rSZ and kSZ.
\begin{figure*}[]
\centering
\includegraphics[width=0.85\textwidth]{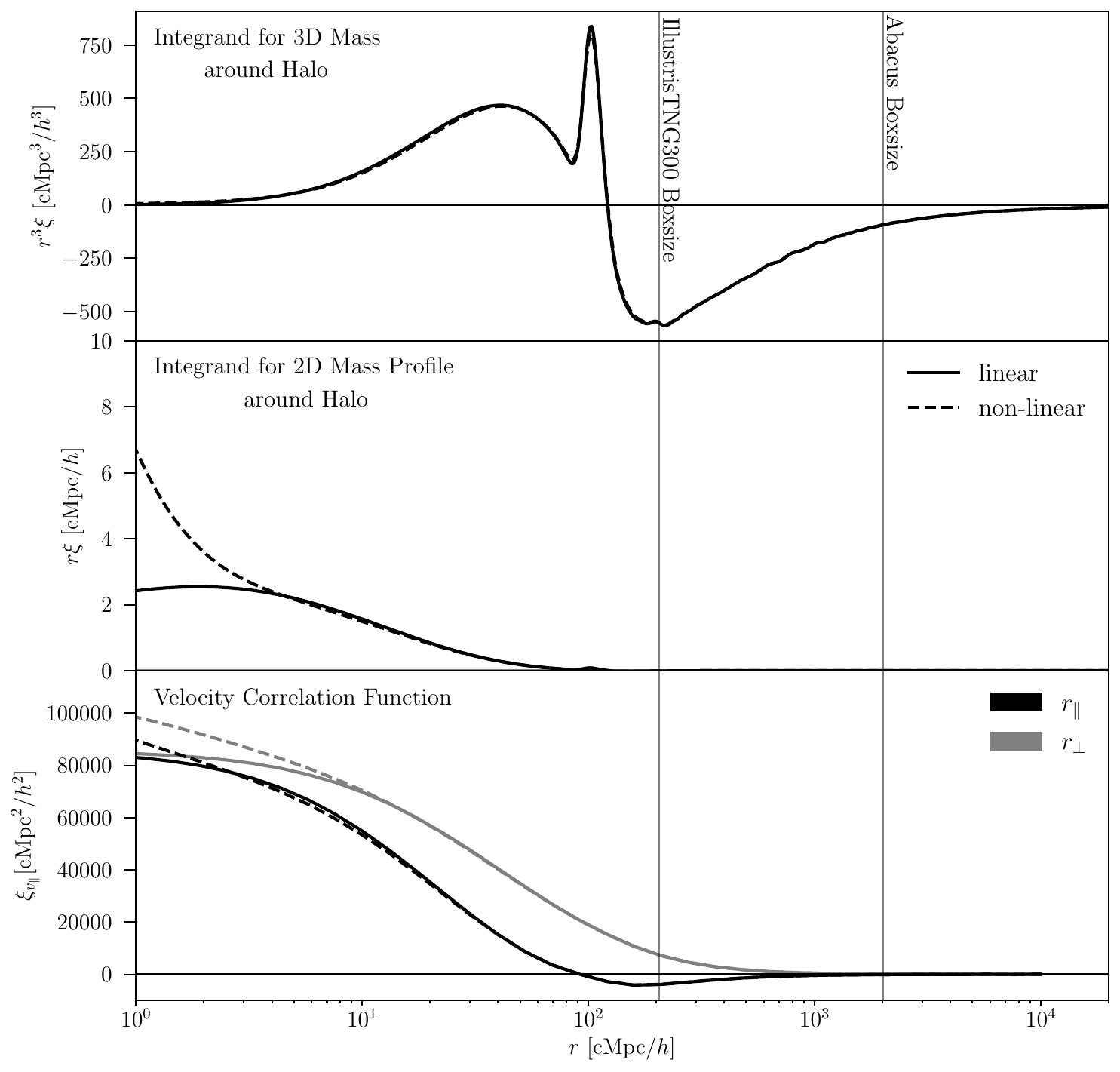}
    \caption{
    From top to bottom, the panels illustrate tests for the convergence scales of the 2-halo term contributions to the 3D and 2D matter distributions, as well as to velocities.
    The curves are plotted such that the integral corresponds to the area under the curve, accounting for the logarithmic scaling of the radius.
    \textit{Top panel:}
    When calculated using 3D spherical shells, the mass contribution from the two-halo term, computed analytically using both the linear (continuous) and nonlinear (dashed) correlation functions, exhibits poor convergence, with significant contributions extending beyond approximately $\sim 5 $Gpc/$h$.
    \textit{Middle panel:} 
    In contrast, the 2D version of this mass term, calculated using cylindrical shells, converges at much smaller scales, around 150~cMpc/$h$ (cMpc stands for comoving Mpc).  
    This indicates that the box size of the IllustrisTNG-300 simulation is sufficient for studies requiring convergence of projected mass profiles.
    \textit{Lower panel:} 
    For analyses involving the kSZ effect, the velocity correlation function, shown for both linear and non-linear theory (black), does not fully converge within the volume of IllustrisTNG-300 but does so in larger simulations such as AbacusSummit.
    This convergence is critical for constraining the contribution of the Doppler term to the kSZ signal, as $\xi_{vv}$ serves as the integrand for the Doppler component (see Eq.~\ref{eq:kSZ_doppler_other} and note the constant mean number density).
    }
\label{fig:linear_correlation_functions}
\end{figure*}
\begin{figure*}[]
    \centering
    \includegraphics[width=1.\textwidth]{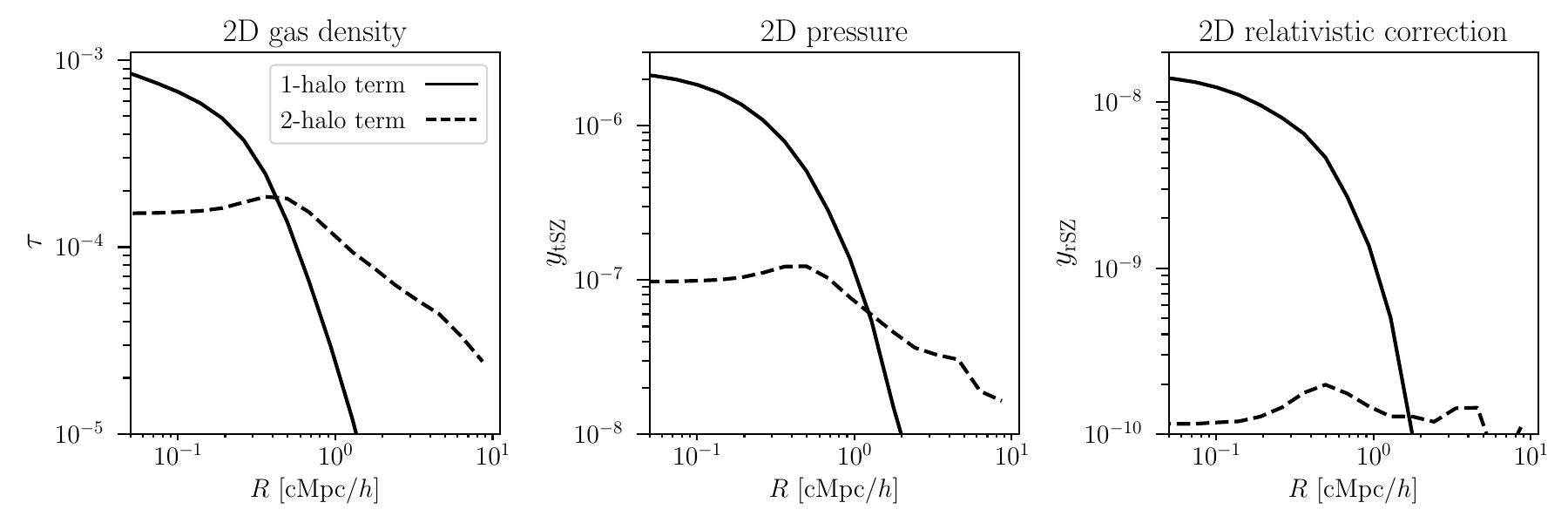}
    \caption{ 
    From left to right, profiles of projected gas density, projected pressure, and projected relativistic corrections to the pressure as a function of radius, as defined in Eqs.~\ref{eq:kSZ},~\ref{eq:compton_tSZ} and~\ref{eq:compton_rSZ}, respectively. 
    These profiles are separated into the 1-halo term (solid line), which dominates at smaller scales ($\lesssim$ 1 cMpc$/h$), and the projected 2-halo term (dashed line), which dominates at larger scales.
    From left to right, as the mass dependence increases ($M$, $M^{5/3}$, and $M^{7/3}$, respectively), we observe a decreasing contribution from the 2-halo term, as the projected profile becomes increasingly dominated by massive halos from neighboring clusters. 
    This trend is accompanied by increasing noise, which is further exaggerated by the logarithmic scale of the plots.
    In the specific case of the projected relativistic SZ, the contribution of the 2-halo term is on the order of a few percent.
    }
    \label{fig:2d_summary}
\end{figure*}

To this end, we aim to identify the scale beyond which the matter contribution from the 2-halo term becomes negligible. 
This can be achieved by analyzing the convergence of the 3D mass density profile within a defined volume surrounding the halo:
\begin{align}
    \int \rho^{\rm 3D}(r) dV & = \int  \rho^{\rm 3D}(r) r^2 \hspace{0.1 cm} dr \hspace{0.1 cm} d\Omega \nonumber \\
    & \propto \int \rho^{\rm 3D}(r) r^3 \hspace{0.1 cm} d\ln{r} \label{eq:corr_3D} \\
    & \propto \int \xi(r) r^3 \hspace{0.1 cm} d\ln{r} \nonumber,
\end{align}
where $\rho^{\rm 3D}(r) = \overline{\rho} (1 + b_{\rm h} \hspace{0.05 cm} \xi(r))$, $\overline{\rho}$ is the mean density, $b_{\rm h}$ is the linear halo bias and $\xi(r)$ is the matter two-point correlation function.
This approximation is accurate on large scales, specifically, where the 2-halo term dominates.

In the top panel of Fig.~\ref{fig:linear_correlation_functions} we show the integrand from Eqs.~\ref{eq:corr_3D} using the linear matter correlation function as a template (continuous line) and the actual measured non-linear matter density profile (dashed line).
The correlation function is computed with the public Python package Nbodykit\footnote{\url{https://nbodykit.readthedocs.io/}} \cite{Hand2018}.
Thus fully capturing the total mass contribution from the 2-halo term in the 3D case, would require probing extremely large simulation boxes, extending to several Gpc/$h$.
Additionally, Fig.~\ref{fig:linear_correlation_functions} illustrates the pronounced impact of the baryon acoustic oscillation (BAO) feature on the distribution of total mass around halos. 
However, this does not affect our analysis, as we are primarily concerned with obtaining converged 2D projected profiles.

If we project the previously defined field along the line of sight ($r_{\|}$), center it on a representative halo, and similarly integrate within a cylindrical volume to determine the convergence scale of the 2D projected matter profile, $\rho^{\rm 2D}(r_{\perp})$, we find that:
\begin{align}
     \rho^{\rm 2D}(r_{\rm \perp}) & = \int \rho^{\rm 3D}(r_{\rm \|}) \hspace{0.1 cm} d(r_{\rm \|}) \nonumber \\
     & = \int \rho^{\rm 3D}(r_{\rm \|}) r_{\rm \|} \hspace{0.1 cm} d\ln{(r_{\rm \|})} \label{eq:corr_2D} \\
     &\propto \int \xi(r_{\|}) r_{\|} \hspace{0.1 cm} d\ln{(r_{\rm \|})} \nonumber,
\end{align}
where $r_{\perp}$ corresponds to the radial distance across the LOS, and the integral extends along the length of the simulation box.

In the middle panel of  Fig.~\ref{fig:linear_correlation_functions} we show the integrand from \ref{eq:corr_2D}.
Thus, the 2D mass profile converges at significantly smaller scales, approximately $\sim$150 cMpc/$h$.
Consequently, we expect the box size of the IllustrisTNG-300 simulation indicated by a vertical line to be sufficiently large to encompass the converged 2D profiles of total or gas mass.

As regards to the specific SZ profiles, the tSZ and rSZ profiles are even more centrally dominated due to their steep scaling with halo mass ($\int y_{\rm tSZ} \hspace{0.1 cm} d^2 \theta \propto M_{\rm h}^{5/3}$ and $\int y_{\rm rSZ} \hspace{0.1 cm} d^2 \theta \propto M_{\rm h}^{7/3}$). 
As a result, these profiles should converge at even smaller scales, meaning that the 2D-projected fields should already be converged at the length of the IllustrisTNG-300 box size.

For the kSZ profile of these halos, we must account for the fact that the velocity field is correlated on large scales \cite{Dam_2021}, resulting in a contribution from the Doppler term, as shown in Sec.~\ref{sec:theory}.
To isolate the kSZ contribution from the halo itself (i.e., the Ostriker–Vishniac term), it is necessary to probe scales at which the Doppler contribution becomes subdominant.
As shown in Eq.~\ref{eq:doppler_harmonic_appendix} in App.~\ref{app:vel_corr_doppler}, the Doppler contribution is integrated along the comoving distance.
We compute the linear velocity correlation function from the linear matter power spectrum, shown in the bottom panel of Fig.~\ref{fig:linear_correlation_functions}.
The velocity correlation function extends to the Gpc/$h$ scale, roughly 5 times the size of the IllustrisTNG-300 simulation box, with an excess of about 10\% of its maximum value.
This correlation extends to Gpc/$h$ scales, roughly ten times the size of the IllustrisTNG-300 box, with an excess of about 10$\%$ of its maximum value. 
Consequently, the Doppler term in IllustrisTNG-300 contributes a nearly constant offset to the kSZ signal, which must be subtracted to recover the halo gas overdensities.

\begin{figure}[]
    \centering
    \includegraphics[width=.93\columnwidth]{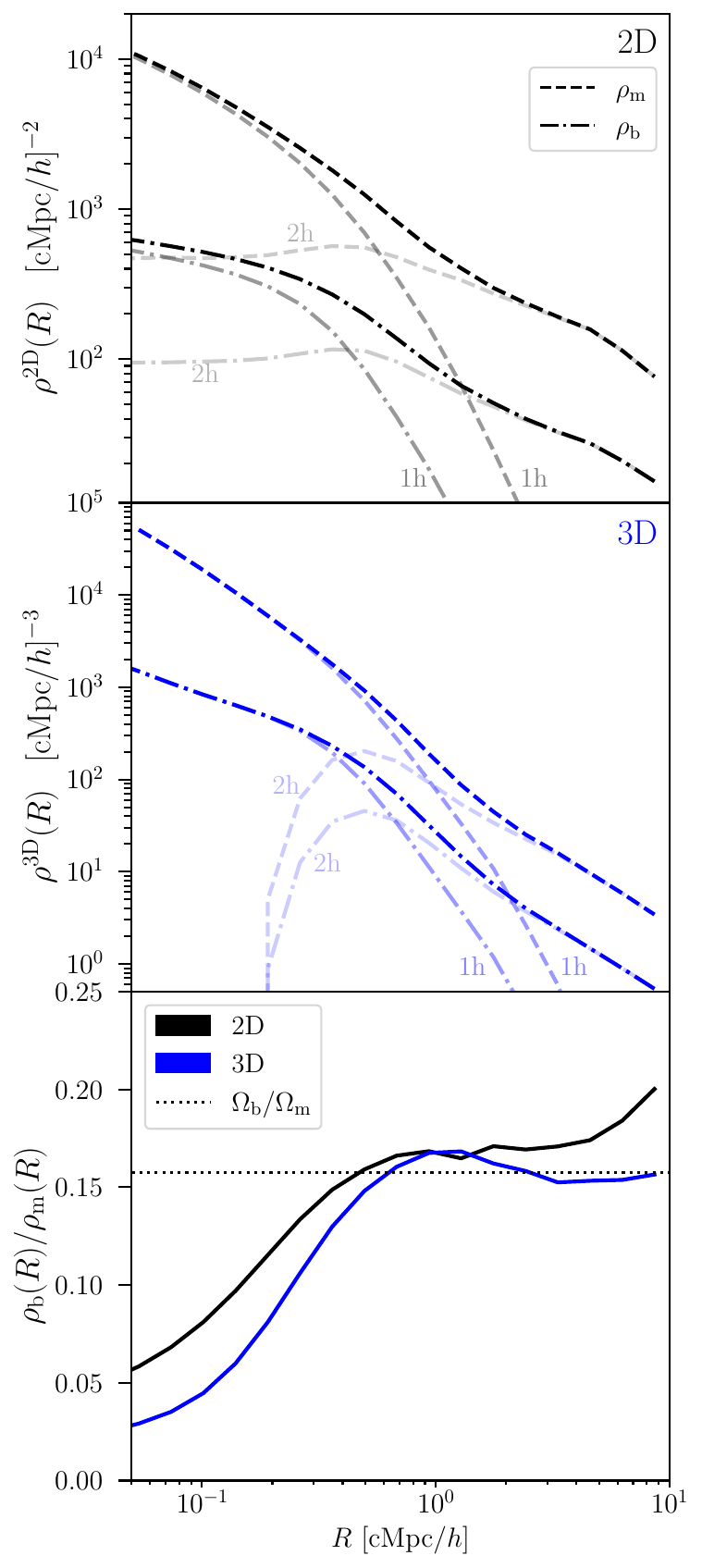}
    \caption{
    The first two panels show the 2D projected and 3D density profiles: baryons are plotted with dash-dotted lines, and total matter (baryons + dark matter) with dashed lines. The upper panel presents the 2D profiles, and the middle panel the 3D ones, with the corresponding 1-halo and 2-halo terms shown faintly. The lower panel displays the ratio of gas to total matter density for both 2D (black) and 3D (blue) profiles. While the 3D ratio converges to the expected value of $\Omega_{\rm b}/\Omega_{\rm m}$ (indicated by the dotted black line), the 2D ratio does not and instead overestimates it, illustrating that gas fractions may not be trivially inferred from projected profiles.
    }
    \label{fig:DM_vs_gas_ratio}
\end{figure}

To ensure the large-scale velocity field contributing to the kSZ effect in Sec.~\ref{sec:results} has converged, we use the larger $N$-body simulation AbacusSummit \cite{Abacus_Maksimova, Abacus_Garrison, Abacus_Hadzhiyska, Abacus_Bose, Abacus_Sandy}. 
This simulation extends to 2000 cMpc/$h$ and includes 150 boxes modeling various cosmological models. 
In the bottom panel of Fig.~\ref{fig:linear_correlation_functions}, it is shown that the size of the Abacus box is sufficiently large for the velocity correlation field to have converged, as indicated by the vertical line.
Since AbacusSummit does not provide gas profiles, we only use them to study the Doppler term in Sec.~\ref{sec:doppler}. 

To select halos in AbacusSummit, we perform abundance matching to emulate the DESI LRG sample by selecting the most massive halos in the simulation down to the mean halo mass of the DESI-like IllustrisTNG sample, ensuring the LRG halo number density is $5 \times 10^{-4}$. 
This results in approximately 4 million halos with masses spanning $5 \times 10^{12} - 2 \times 10^{15} M_{\odot}~h^{-1}$. 
To speed up our numerical computations, we select a random subsample of 100,000 halos for stacking.

\section{Results}
\label{sec:results}

\subsection{Projected 2D profiles around simulated DESI LRGs}

The thermodynamic 2D profiles derived from a color-selected LRG sample from IllustrisTNG are shown in Fig.~\ref{fig:2d_summary}. 
Three projected profiles are measured: gas density, pressure and relativistic SZ. 
Each profile incorporates both the 1-halo and 2-halo terms.
The 3D version of these same profiles is shown in Fig.~\ref{fig:3d_summary} from App.~\ref{sec:3D_profiles}.

We corroborate the findings from \cite{Moser2021}:
in all cases, the 1-halo term dominates the profiles in the range $\lesssim$0.3–2~cMpc$/h$. 
This finding is also observed in the 3D profiles.

A fundamental difference between the 2D and 3D profiles is the plateau shape of the 2-halo term at scales where the 1-halo term dominates in the 2D case study.
This suggests that at any radius, there will be a non-negligible contribution from neighboring halos, most importantly for the density: the 2-halo contribution is approximately 20\% of the 1-halo contribution at $R \approx 0.1$~cMpc/$h$.
For the tSZ signal, we find a similar trend to that reported by \cite{Li2011}, where the contribution from the two-halo term can reach up to $\sim 40\%$ for group-scale halos with masses around $10^{13} M_{\odot}/h$.
We conclude that, for kSZ, the 2-halo term remains non-negligible even at small radii where the 1-halo term dominates, in contrast to the 3D profiles, where the 2-halo contribution is negligible at these scales.
This conclusion changes once a compensated aperture photometry (CAP) filter is applied, as flat profiles are removed out (see Sec.~\ref{sec:doppler}).

\begin{figure*}[]
    \centering
    \includegraphics[width=1.\textwidth]{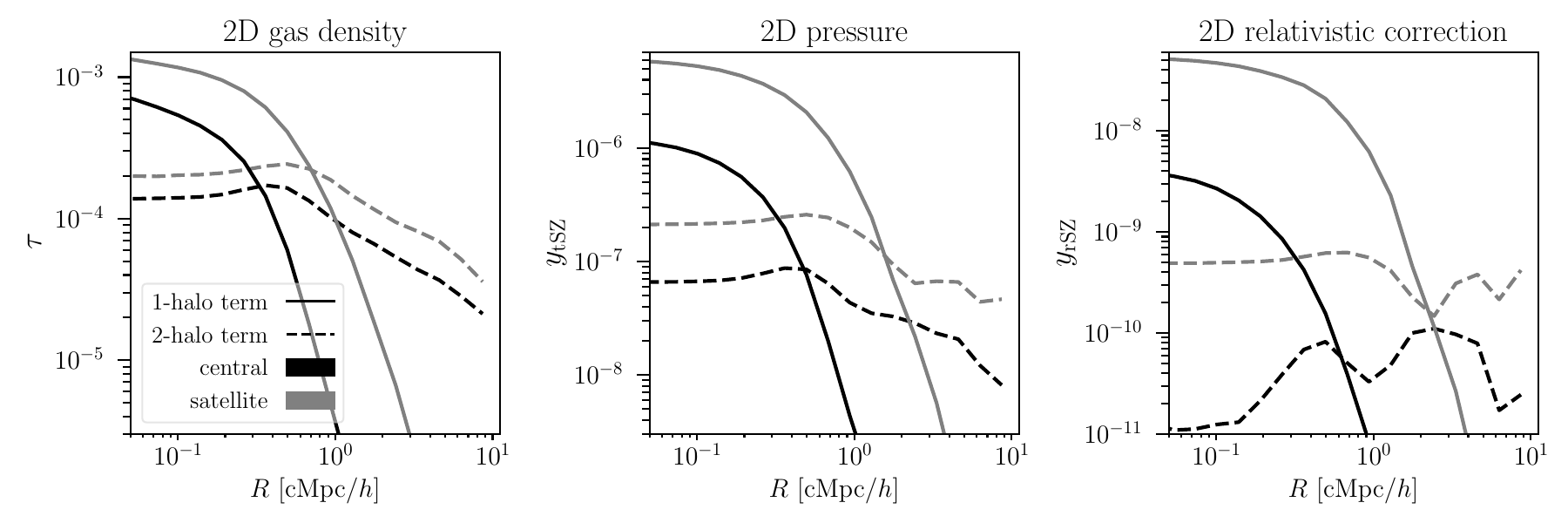}
    \caption{
    Similar to Fig.~\ref{fig:2d_summary}, except this figure splits the LRG sample into central (black) and satellite (grey) galaxies. 
    Satellites tend to reside in larger halos, which have higher mass and are surrounded by more massive neighbors. Their 1-halo and 2-halo terms are thus larger than for centrals.
    This effect is especially pronounced in pressure and projected relativistic SZ, as these quantities have an even steeper mass dependence. 
    However, given the total number of satellites is lower than the centrals, their contribution to the average profiles for centrals plus satellites is reduced by their fraction compared to the curves shown.
    }
    \label{fig:sat_vs_cent_summary}
\end{figure*}

Furthermore, when comparing from the left to the right panel, the fraction of the total distribution accounted for by the 2-halo term becomes smaller. 
Although it is a significant contribution to gas density at any scale, it is on the order of one percent for rSZ, making it practically negligible. 
This behavior is expected, as discussed in Sec.~\ref{sec:theory}, since the mass dependence weakens from the 1-halo term to the 2-halo term due to the halo bias.
We also note that the 2-halo term of both the pressure and the relativistic SZ becomes noisier at large radii due to the high concentration of particles and the low gas density in the outskirts of halos.
Despite this, our argument remains valid, as it is evident that the overall 2-halo term decreases when the mass dependence is stronger.

\subsection{Possible overestimation of gas fractions from 2D projected profiles}

A commonly quoted quantity in gas measurements is the baryon fraction within a given radius of a halo. 
For 2D observables such as SZ profiles, projection effects can bias the inferred gas content within a particular aperture. 
In this section, we investigate whether the baryon fraction can be reliably estimated from 2D projected profiles of matter and gas density.

We calculate the ratio of baryons to total matter for both the 2D projected and 3D profiles as a function of radius and show their differences in Fig.~\ref{fig:DM_vs_gas_ratio}.
As the radius increases, the 3D profile converges to the cosmologically expected baryon-to-matter ratio, $\Omega_{\rm b}/\Omega_{m}$, as anticipated. 
However, the 2D density profile tends to overestimate the gas fraction, as it integrates the total gas content along a cylinder spanning the length of the simulation box, thereby enhancing the observed profiles.

This result is particularly noteworthy because it demonstrates that
inferring true baryon fractions from the large-scale limit of 2D projected profiles may not be trivial.
Recently, \cite{hadzhiyska2025missingbaryonsrecoveredmeasurement} derived the baryon fraction by stacking kSZ and lensing profiles. 
While the results are limited by large, highly correlated error bars, primarily due to the impact of CMB primary anisotropies at large scales, upcoming data from observatories such as Rubin and SO will reduce statistical uncertainties, making it increasingly important to account for potential projection effects.
Additional filtering such as CAP may modify this discussion.

\subsection{
Satellite fraction and 2D profile sensitivity
}

Satellite galaxies, which typically reside in massive halos and are often offset from halo centers, significantly influence SZ profiles, as shown in \cite{mccarthy2024flamingocombiningkineticsz}.
In Fig.~\ref{fig:sat_vs_cent_summary}, we extend the analysis presented in Fig.~\ref{fig:2d_summary} by separating the galaxy sample into central and satellite galaxies and plotting the average profile for each type.
To compute the overall average for the full sample (centrals plus satellites), these curves must be weighted by the factors $f_{\rm cen} = 1 - f_{\rm sat}$ and $f_{\rm sat}$, respectively.
\begin{figure*}[]
    \centering
    \includegraphics[width=0.98\textwidth]{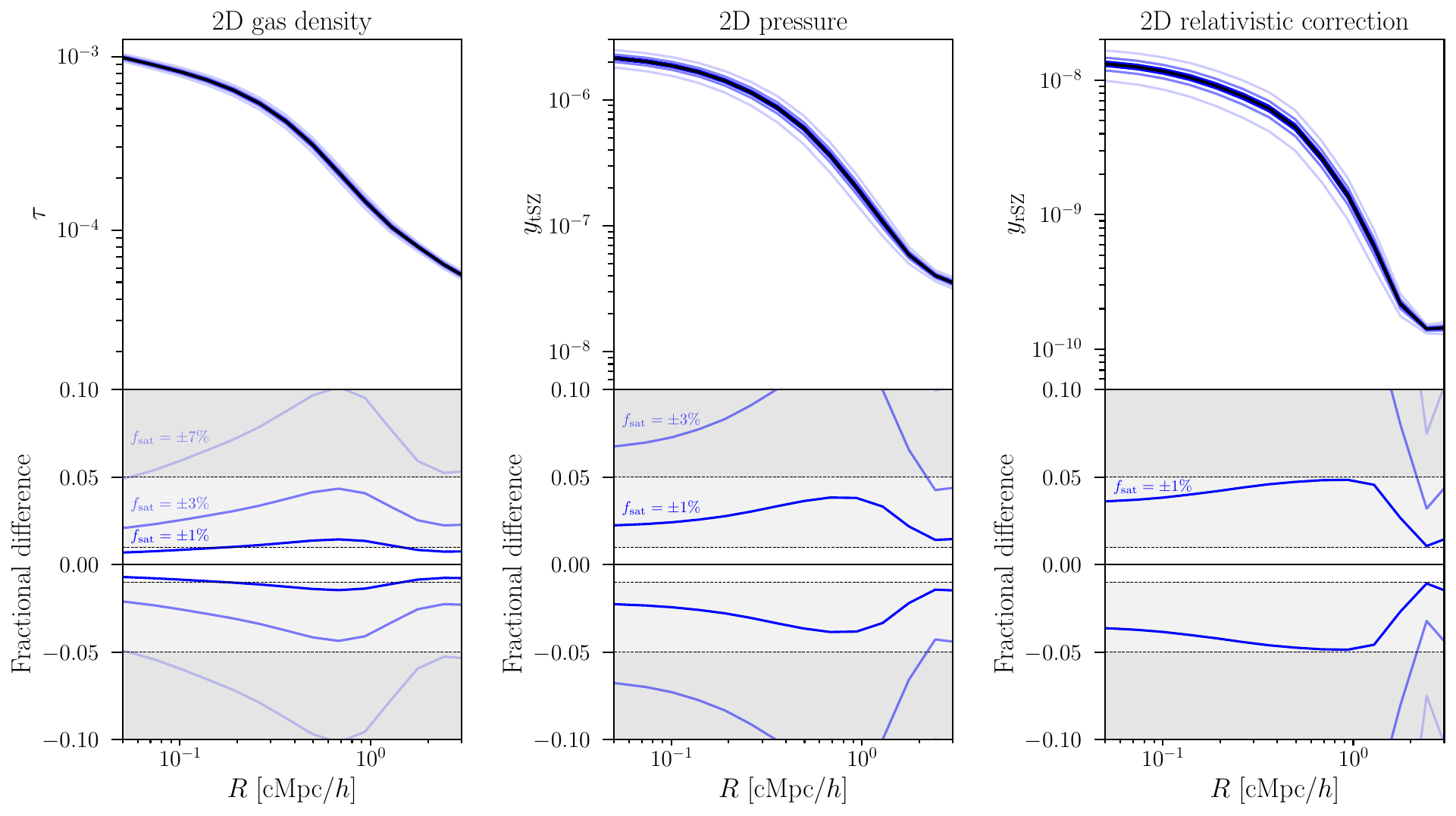}
    \caption{
    The top panels show the projected 2D profiles of gas density, pressure, and relativistic SZ for varying satellite fractions, and their fractional errors within a $\pm1\%$ (white), $\pm5\%$ (light gray) and $\pm10\%$ (gray) bands in the lower panels.
    Each profile represents a linear combination of the central and satellite profiles from Fig.~\ref{fig:sat_vs_cent_summary}, with a satellite fraction of 20$\% \pm 1\%$, $20\pm 3\%$ and $20\pm 7\%$.
    The fiducial satellite fraction, denoted as $f_{\rm sat}$, corresponds to 20$\%$.
    From left to right, a $\pm$1$\%$ change in the fiducial satellite fraction results in an increasing fractional error, reaching up to 2$\%$ for the projected gas density, $\sim 4 \%$ for the projected pressure, and 5$\%$ for the relativistic SZ profile.
    This allows us to quantify the maximum allowable uncertainty in $f_{\rm sat}$ to ensure that its propagated error to the SZ profiles (shape and amplitude) remains constrained by certain percentage.
    }
    \label{fig:sat_frac_slope}
\end{figure*}

As shown in \cite{popik2025impactshalomodelimplementations}, we confirm that satellites exhibit larger 1-halo terms than centrals, as they are more likely to reside in massive halos. 
This is also shown in \cite{Li2011}, where the $y_{\rm tSZ}$ 1-halo term of the satellites dominates in group-size halos.
Consequently, these massive halos have higher clustering bias, resulting in a larger 2-halo term, which is also evident in the 2D profiles.
However, given the total number of satellites is lower than the centrals, their individual contribution to the total profile is in practice down-weighted by their fraction.

Given the substantial differences between satellite and central galaxies, we investigate how the total profiles (combining 1-halo and 2-halo terms) respond to variations in the satellite fraction.
Specifically, we vary the fiducial satellite fraction,  $f_{\rm sat} = 20\%$, by $\pm 1\%, \pm 3\%$ and $\pm 7\%$.
For reference, the 1-$\sigma$ uncertainty on the satellite fraction from DESI LRG galaxies, based on the first HOD analysis, is approximately  $\pm 1\%$ \cite{yuan2023desi}, although we note that this analysis may not fully marginalize over physical models of assembly bias.
The resulting profiles are shown in Fig.~\ref{fig:sat_frac_slope}.

As expected, effects with the steepest mass dependence exhibit the largest response to changes in the satellite fraction. 
Specifically, a $\pm1\%$ variation in $f_{\rm sat}$ results in approximately a 2\% change in the amplitude and shape of the 2D gas density profiles, a 4\% change for the 2D pressure, and a 5\% change for the 2D relativistic SZ, respectively.
This highlights the importance of accurately constraining the satellite fraction, particularly for quantities with strong mass dependence such as the pressure and relativistic SZ profiles. 
By quantifying this sensitivity, we can determine the maximum allowable uncertainty in $f_{\rm sat}$ to ensure that its propagated error in the SZ profiles remains within acceptable limits.
For example, achieving a $\pm 1\%$ precision on gas profiles derived from kSZ studies, requires that the uncertainty on $f_{\rm sat}$ remain sub-percent.

\begin{figure*}[]
    \centering
    \begin{subfigure}[b]{0.60\textwidth}
        \includegraphics[width=0.95\textwidth]{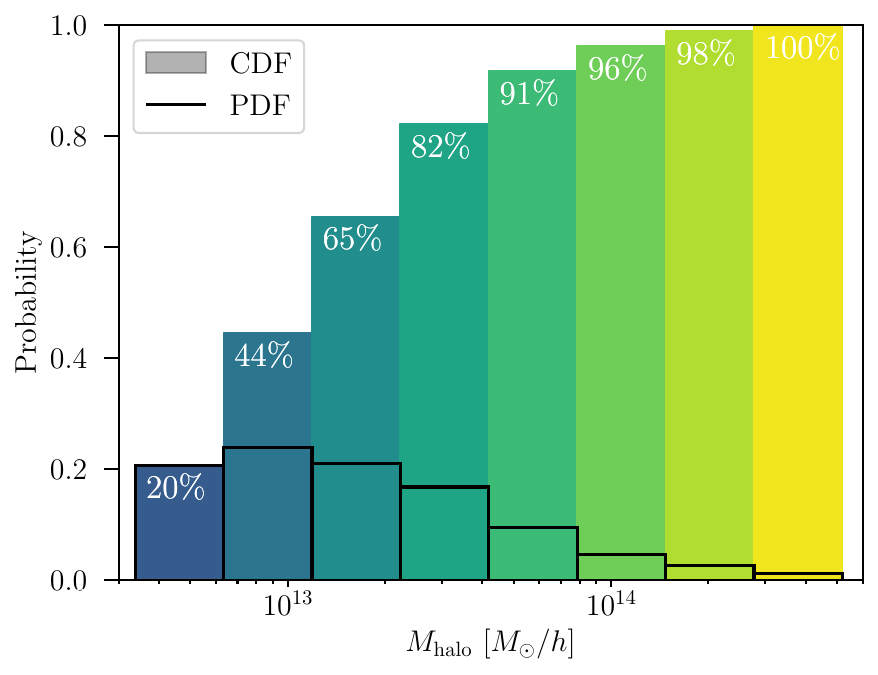}
    \end{subfigure}
    \begin{subfigure}[b]{0.99\textwidth}
        \includegraphics[width=0.99\textwidth]{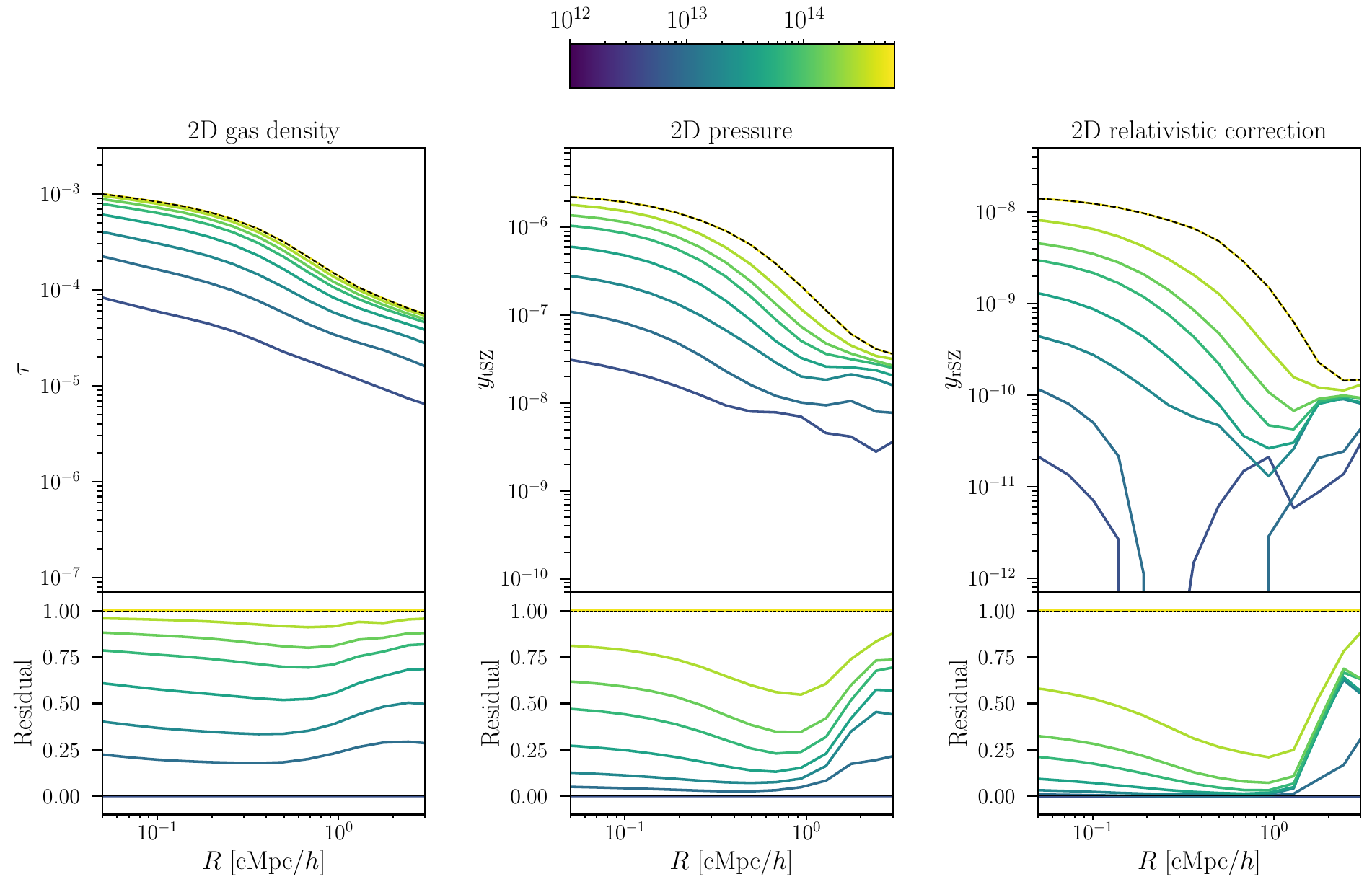}
    \end{subfigure}
    \caption{
    The upper panel shows the probability density and cumulative halo mass distribution of the DESI-like LRG sample from IllustrisTNG.
    The lower panel presents a plot similar to Fig.~\ref{fig:sat_frac_slope}, but with the sample now split by cumulative halo mass.
    From left to right, increasing cumulative mass results in profiles that get closer to the fiducial one shown as a black dashed line.
    It is also evident that stronger mass dependence causes more of the curves to fall below the fiducial profile in the fractional profiles, indicating that the most massive halos pull the average profile upward substantially.}
    \label{fig:mass_buildup}
\end{figure*}

\subsection{
Impact of the most massive host halos in the LRG galaxy sample
}

The simulated LRG sample spans host halo masses from approximately $3 \times 10^{12}$ to $5 \times 10^{14} M_{\odot}/h$, as shown in the upper panel of Fig.~\ref{fig:mass_buildup}.
Given our earlier discussion when separating centrals and satellites, we anticipate that the most massive halos will dominate the profiles, particularly for the relativistic SZ signal.
To test this, we sort the LRG sample by host halo mass and evaluate their cumulative contributions, shown in the lower panel of Fig.~\ref{fig:mass_buildup}.

The upper panel shows that group-scale halos (around $M_{\rm h} \approx 10^{13} M_{\odot}/h$) are by far the most numerous. 
However, the lower panel reveals that the most massive halos, despite constituting only a small fraction of the population, can dominate the overall signal.
This effect is most pronounced in the case of the relativistic SZ, where nearly the entire contribution arises from the most massive halos.
Notably, for the tSZ and rSZ signals, excluding the most massive halos has a dramatic impact: 
removing the top 35\% and 2\% of halos in mass reduces the average profile amplitude approximately up to 75\%, respectively.
For the kSZ, masking only the most massive 2\% of halos already lowers the signal amplitude to about 90\% of its original value, which, given the current level of precision in ongoing measurements, represents a significant source of systematic uncertainty that must be carefully accounted for in analyses.

Moreover, because the contribution from this high-mass tail increases for observables with a steeper mass dependence, the average gas density, pressure, and relativistic SZ effectively probe different subsets of the halo population, even when measured around the same galaxy sample.
As demonstrated in \cite{Li2011}, mass-assignment errors exacerbate this effect, since a larger mass variance within the sample can bias the inferred SZ signals.
Additionally, we confirm the results from \cite{Hill_2018} regarding the relatively larger importance of the 2-halo term in the lower mass regime in the 2D pressure profiles (middle panel).

\subsection{
Suppression of the artificial Doppler term by the CAP filter: implications for kSZ measurements
}
\label{sec:doppler}

\begin{figure}[]
    \centering
    \includegraphics[width=\columnwidth]{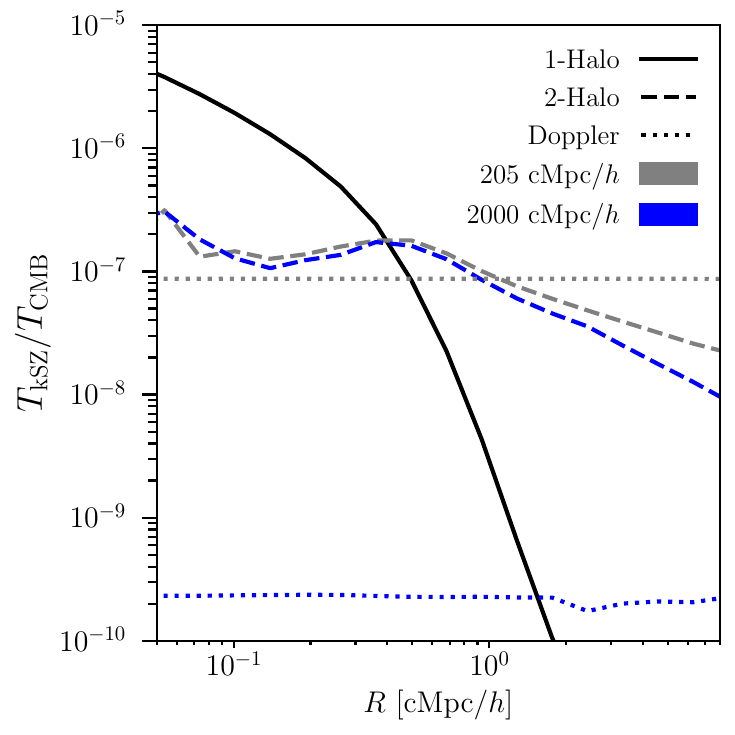}
    \caption{
    The projected Doppler terms for IllustrisTNG (in gray) and AbacusSummit (in blue) are shown here as flat dotted lines. 
    The kSZ temperature shift of the 1-halo and 2-halo terms, excluding the Doppler contribution, are displayed as solid and dashed lines, respectively.
    We have used $v_{\rm \parallel} = v_{\parallel}^{\rm RMS}$ from the simulations. 
    As anticipated from Fig.~\ref{fig:linear_correlation_functions}, as the cylinder length increases, the contribution from the Doppler term decreases.
    }
    \label{fig:stacked_kSZ} 
\end{figure}
\begin{figure}[]
    \centering
    \includegraphics[width=\columnwidth]{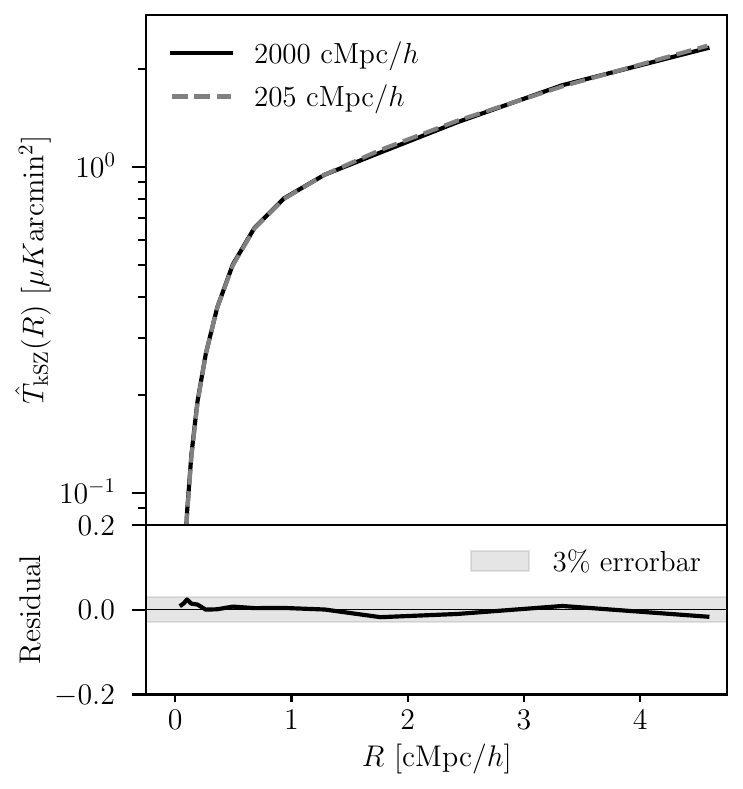}
    \caption{
    The CAP-filtered kSZ profile was computed for IllustrisTNG (denoted by 205 cMpc/$h$ in dashed gray) and AbacusSummit (denoted by 2000 cMpc/$h$ in solid black). 
    As the cylinder length increases, the profile converges to a well-defined curve, as shown in the residual plot below.
    }
    \label{fig:CAP_filter}
\end{figure}

The Doppler term can dominate kSZ measurements if the survey is not sufficiently deep along the line of sight, thereby hindering a robust separation of the halo gas content.
As shown in Fig.~\ref{fig:linear_correlation_functions}, the velocity correlation function remains significant at the scale of the IllustrisTNG box size, implying a non-negligible Doppler contribution as described in Eq.~\ref{eq:doppler_harmonic_appendix}, while it becomes negligible at the Abacus box size.
We therefore investigate whether the integration length along the LOS is indeed sufficient for positive and negative velocity modes to cancel effectively, thereby suppressing the Doppler term to a subdominant level.

In Fig.~\ref{fig:stacked_kSZ}, we present the 1- and 2-halo contributions to the projected kSZ profiles, integrated along the LOS over the lengths of IllustrisTNG (205 cMpc/$h$) and AbacusSummit (2000 cMpc/$h$), together with the corresponding Doppler terms.
For the IllustrisTNG box size, the Doppler term dominates at scales $\gtrsim 1$ cMpc/$h$. 
However, when integrated over the larger AbacusSummit box length, the velocity modes effectively cancel, leading to a suppression of the Doppler contribution by nearly three orders of magnitude.

Although the Doppler term may still be relevant in hydrodynamical simulations such as IllustrisTNG, it is effectively mitigated when a CAP filter from Eq.~\ref{eq:CAP} is applied. 
The CAP filter removes contributions from a constant field at any given radius, thereby automatically suppressing the Doppler term.

Fig.~\ref{fig:CAP_filter} further illustrates this behavior: 
as the cylinder length increases from the IllustrisTNG box size to that of AbacusSummit, the CAP filter converges to a well-defined shape, with residual fluctuations remaining below 3$\%$. 
This behavior is also observed in \cite{Moser2023}, where the line-of-sight length is varied between 10–100 cMpc.
As a result, the kSZ profiles in analyses such as \cite{hadzhiyska2024evidencelargebaryonicfeedback, riedguachalla2025backlightingextendedgashalos} provide an accurate representation of the 1- and 2-halo terms, without contamination from the Doppler contribution.
We note that that the CAP suppression demonstration is for idealized conditions (e.g. true velocities, periodic box, single snapshot) and that residual Doppler contamination in real surveys depends on survey depth, geometry, and velocity reconstruction noise.

\section{Conclusions}
\label{sec:conclusions}

In this work, we have studied the projected thermodynamic profiles associated with the thermal, relativistic, and kinematic SZ effects using a simulated DESI-like sample of LRGs from the IllustrisTNG simulation. 
Rather than treating the hydrodynamical simulation as a definitive model of gas physics, we use it as a representative example with a realistic level of complexity, in order to get a sense for the rough size of various modeling effects.
Aditionally, we defer light cone effects, beams, noise, and component separation for future studies.
By decomposing the SZ contributions into 1-halo and 2-halo terms and analyzing them in both 2D and 3D, we quantify several key sources of bias and signal dependence that are directly relevant to interpreting observational SZ measurements.

We find that the 1-halo term dominates the SZ profiles at small scales ($\leq$ 2 cMpc/$h$), while the 2-halo term, although subdominant, is non-negligible even at small radii due to projection effects. 
This is particularly important for projected gas density, where the 2-halo contribution remains significant ($\sim 20\%$ of the 1-halo term at 0.1 cMpc/$h$). 
We also show that 2D projections systematically overestimate the baryon fraction at large radii compared to 3D profiles, introducing biases that could affect gas fraction measurements from stacked SZ observations.

A central finding is that the SZ profiles around satellite galaxies are significantly larger than those around centrals, owing to their preferential location in massive halos. 
However, because satellites are less numerous, their contribution is down-weighted in the stacked signal. 
As a result, the total SZ profile represents a weighted average between the central and satellite contributions, that must be taken into account in the profile analyses.
Specifically, the 1-halo and 2-halo terms for satellites are substantially larger than for the pressure and relativistic SZ profiles, both of which are strongly mass-dependent. 
Consequently, the SZ profiles are highly sensitive to the satellite fraction: even a $\pm 1 \%$ uncertainty in the satellite fraction can lead to a 2–5$\%$ change in the SZ signal amplitudes.

Further, we demonstrate that the most massive halos, although comprising only a small fraction of the population, disproportionately shape the stacked SZ profiles. 
For example, the relativistic thermal SZ correction is almost entirely sourced by the top 2$\%$ most massive halos, and removing them reduces the signal's amplitude by $\sim 75\%$. 
This underscores that different SZ observables probe distinct halo populations, even when measured over the same galaxy sample.

Finally, we analyze the Doppler contamination in the kSZ profiles and find that it can be significant for small simulation volumes like IllustrisTNG, where it is comparable to the 2-halo term. 
However, this contribution is strongly suppressed with increasing LOS integration length, and more importantly, it is effectively removed by the CAP filtering technique used in real observational analyses.

Looking ahead, several natural extensions of this work can be explored. 
One such avenue is investigating the kSZ profile as a function of e.g. satellite fraction and halo mass, particularly examining the contributions from the 1-halo and 2-halo terms. 
Additionally, applying these methods to lightcone simulations would more accurately capture observational conditions, including redshift evolution and survey geometry, which could have important implications for the Doppler term.
Incorporating miscentering effects, as discussed in \cite{popik2025impactshalomodelimplementations}, could further enhance the precision of the SZ profiles by accounting for offsets in the halo center. 
Given the strong sensitivity to satellite fractions, careful investigation of the impact of assembly bias is also warranted.

Overall, our results highlight the critical role of halo mass distribution, satellite fraction, and projection effects in interpreting SZ measurements.
These findings are directly relevant for current and future CMB–galaxy cross-correlation studies using DESI, LSST, Simons Observatory, and CMB-S4, where accurate modeling of these effects is essential for extracting robust constraints on baryonic feedback, gas physics, and cosmology.

\section*{Acknowledgments}

We thank Andrew Pontzen, Boryana Hadzhiyska, Elisabeth Krause, Frank Qu, Hiranya Peiris, Nicholas Battaglia, Simone Ferraro, Tim Eifler, and Tom Abel and  for useful discussions.

This work received support from the U.S. Department of Energy under contract number DE-AC02-76SF00515 to SLAC National Accelerator Laboratory and from the Kavli Institute for Particle Astrophysics and Cosmology at Stanford and SLAC.

\begin{figure*}[]
\centering
\includegraphics[width=1.\textwidth]{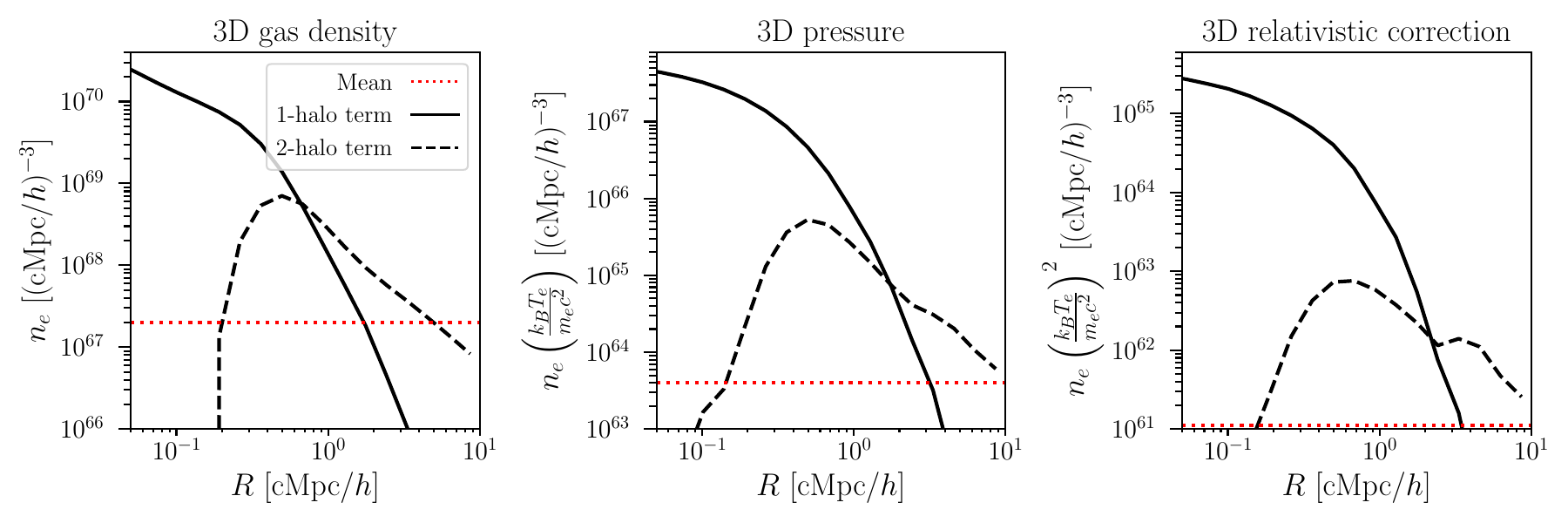}
\caption{
Similar to Fig.~\ref{fig:2d_summary}, but now focusing on the 3D distributions, we find that the 2-halo term becomes relevant only at larger radii. 
The simulation average quantity is shown in red. 
Moreover, its relative contribution diminishes with increasing mass dependence, being less significant for quantities like pressure and the relativistic SZ compared to gas density.
}
\label{fig:3d_summary}
\end{figure*}

\begin{figure*}[]
\centering
\includegraphics[width=1.\textwidth]{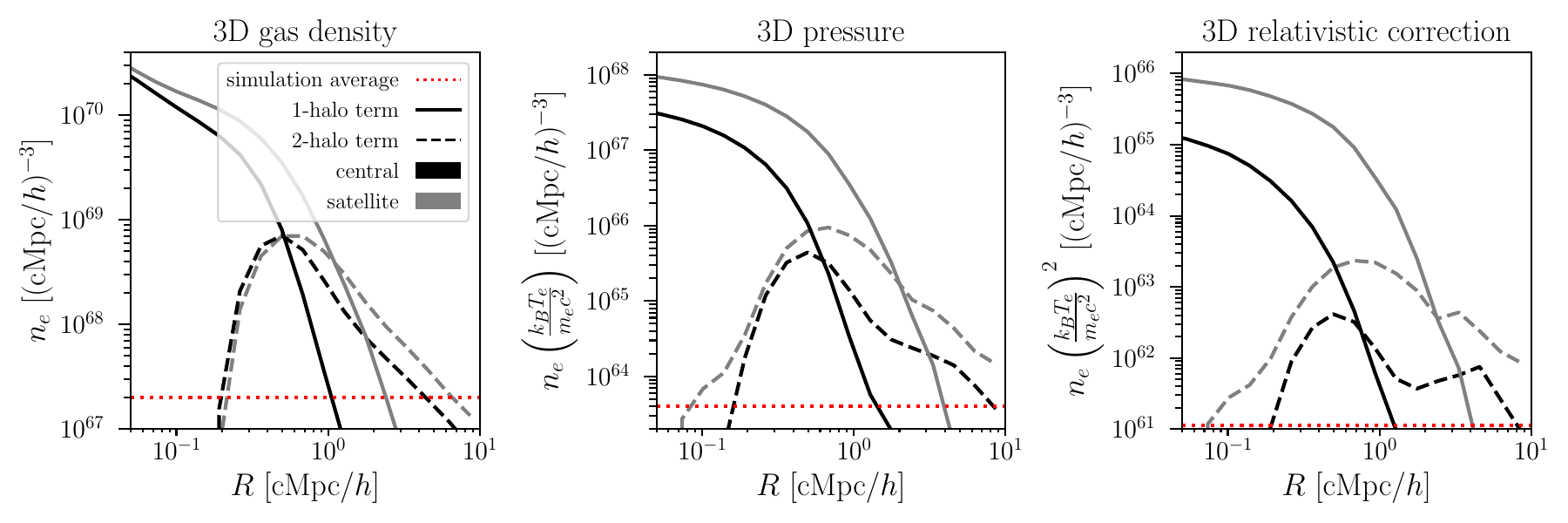}
\caption{
The 3D distributions around satellite (orange) and central (green) subhalos are shown separately. 
For both the 1-halo (dashed) and 2-halo (solid) components, the satellite contribution is consistently larger. 
This discrepancy grows with increasing mass dependence, being most pronounced in quantities like pressure and rSZ.
}
\label{fig:3d_summary_sat_vs_cent}
\end{figure*}

\appendix

\bibliographystyle{prsty.bst}
\bibliography{refs}

\section{Stacked 3D Profiles}
\label{sec:3D_profiles}

Fig.~\ref{fig:3d_summary} shows the 3D thermodynamical profiles introduced in Sec.~\ref{sec:observables}. 
In contrast to the 2D versions presented in Fig.~\ref{fig:2d_summary}, the 3D profiles exhibit similar trends but also reveal clear distinctions, which we outline below.

In both cases, 2D and 3D, the 2-halo term decreases as the dependence on mass of the profile studied, i.e., from 3D gas density to pressure to rSZ, the significance of the 2-halo term in comparison to the 1-halo term decreases.
One notable distinction is that within a small 3D radius, the 2-halo contribution becomes negligible. 
This contrasts with the 2D case, where projection effects along the line of sight amplify the 2-halo signal. 
This difference should be carefully considered when evaluating its significance in the 2D analyses.

Another valuable aspect to examine in 3D is the comparison between the 2-halo term and the simulation-wide average, shown in red in Fig.~\ref{fig:3d_summary}. 
This helps indicate how far out it remains meaningful to discuss the 2-halo contribution, and beyond what radius the measurement simply reflects the mean density of the simulation box. 
As expected, due to their strong mass dependence, this transition occurs at much larger scales for pressure and the rSZ effect than for gas density. 
This is an important consideration when interpreting these profiles at large radii.

Fig.~\ref{fig:3d_summary_sat_vs_cent} presents the breakdown of satellite and central contributions to the 3D distributions. 
As in the 2D case, satellite galaxies, which hosts more massive halos, clearly higher average values, with the separation becoming more pronounced at higher masses. 
This trend is strongest in the relativistic SZ, followed by pressure, and is least apparent in gas density.

\section {Velocity correlation function and Doppler term}
\label{app:vel_corr_doppler}

To calculate the Doppler term in Eq.~\ref{eq:kSZ_doppler_other}, we begin by defining the peculiar velocity field. 
In linear theory, this is given by:
\begin{equation}
    {v_i}(\mathbf{k}) = - i \hspace{0.1 cm} a H f \frac{k_i}{k^2} \delta(\mathbf{k}) 
\end{equation}
where the subscript $i$ stands for the direction of the velocity field, $a$ is the scale factor, $H$ is the Hubble parameter and $f$ is the logarithmic growth rate of structure.

The velocity correlation function can be written as:
\begin{align}
    \xi_{v_i v_j}(\mathbf{r}) 
    &= \left< v_i(\mathbf{x}) v_j(\mathbf{x} + \mathbf{r})  \right> \\
    &= \int \frac{d^3 \mathbf{k}}{(2\pi)^3} \frac{k_i k_j}{k^2} P_{v}(k) e^{-i \mathbf{k} \cdot \mathbf{r}}
    \label{eq:vel_corr}
\end{align}
where $P_v$ is the isotropic velocity power spectrum:
\begin{equation}
    P_{v}(k) = \frac{(a H f)^2}{k^2} P_m(k)
\end{equation}

From Eq.~\ref{eq:vel_corr} we find that the velocity correlation function is an anisotropic field, and can be expressed in terms for two isotropic functions $\xi_{v_{\parallel}}$ and $\xi_{v_{\perp}}$ \cite{Gorski1989}:
\begin{equation}
    \xi_{v_i v_j}(\mathbf{r}) = \xi_{v_{\perp}}(r)\delta^{K}_{ij} + [\xi_{v_{\parallel}}(r) - \xi_{v_{\perp}}(r)] \left( \frac{r_i r_j}{r^2} \right)
\end{equation}
where $\delta^{K}_{ij}$ is the delta Kronecker function.
The isotropic fields are given by:
\begin{align}
    \xi_{v_{\parallel}}(r) &= \int dk \frac{k^2}{2 \pi^2} P_{v}(k) \left[ j_0(kr) - \frac{2 j_1(kr)}{kr} \right] \\
    \xi_{v_{\perp}}(r) &= \int dk \frac{k^2}{2 \pi^2} P_{v}(k) \frac{j_1(kr)}{kr}
    \label{eq:decomposition_vel}
\end{align}
where the functions $j_0$ and $j_1$ are the spherical Bessel functions first kind of order zero and first respectively.
For a thorough analysis of the velocity correlation equations including e.g. redshift space distortions, which we do not explore as our reconstructed velocities corresponds to much larger scales, see \cite{Blake2023}.

As shown in Eq.~\ref{eq:kSZ_doppler_other}, the CMB temperature contribution is given by:
\begin{align}
     \frac{\Delta T_{\rm Doppler}}{T_{\rm CMB}} 
     =& 
     - \int  \overline{n}_e  \frac{v_{\parallel}}{c} \sigma_{\rm T} \ d\ell \\
     =& 
     -\frac{\overline{n}_{e, 0} \sigma_{\rm T}}{c} \int v_{\parallel} (1+z)^2 d\chi \\
     \propto& 
     - \int v_{\parallel} g(z) d\chi .
\end{align}
where we have defined $g(z) = \frac{\overline{n}_{e, 0} \sigma_{\rm T}}{c} (1+z)^2$.

We can derive the corresponding two–point angular correlation function of the Doppler-induced CMB temperature anisotropy:
\begin{equation}
    C^{\rm Doppler} (\theta) \propto
    \int d\chi d\chi^* g(z) g(z^*) \xi_{v_{\parallel}}(r).
\end{equation}

And finally, the angular power spectrum in harmonic space would be given by:
\begin{align}
    C_{\ell}^{\rm Doppler}
    \propto&
    \int_{-1}^1 d\mu \hspace{0.1 cm} P_{\ell}(\mu) \hspace{0.1 cm} C^{\rm Doppler}(\theta)  \\
    =& 
    \int_{-1}^1 d\mu \hspace{0.1 cm} P_{\ell}(\mu) 
    \int d\chi d\chi^* g(z) g(z^*) \xi_{v_{\parallel}(}r)
    \label{eq:doppler_harmonic_appendix}
\end{align}
where $P_{\ell}$ is the Legendre polynomial of order $\ell$.

In reality, we use galaxies as tracers of the velocity field, therefore, we define the cross-correlation between the galaxy overdensity 
\begin{align}
    \delta_g =& \int d\chi \hspace{0.1 cm} n_g(\chi) b(z) \delta_m  \\
    \propto& \int d\chi \hspace{0.1 cm} h(z) \delta_m, 
\end{align}
where we have defined $h(z) = n_g(\chi) b(z)$
and each cross-contribution as:
\begin{align}
    \psi_{i}(\mathbf{r}) 
    &= \left< \delta_g(\mathbf{x}) v_j(\mathbf{x} + \mathbf{r})  \right> \\
    &= \int \frac{d^3 \mathbf{k}}{(2\pi)^3} \frac{k_i}{k} P_{gv}(k) e^{-i \mathbf{k} \cdot \mathbf{r}}
    \label{eq:vel_corr}
\end{align}
where $P_{gv}$ is the isotropic galaxy-velocity cross-power spectrum:
\begin{equation}
    P_{gv}(k) =  i b \frac{a H f}{k} P_m(k)
\end{equation}
where we have neglected non-linear galaxy bias.
Similarly as done in Eq.~\ref{eq:decomposition_vel}, the corresponding cross-correlation for velocities along the LOS is given by:
\begin{equation}
    \psi_{gv_{\parallel}}(r)  = \int dk \frac{k^2}{2 \pi^2} P_{gv}(k) j_1(kr).
\end{equation}

Therefore, the biased cross Doppler-induced term in this case would be
\begin{equation}
    C^{\rm B-Doppler} (\theta) \propto
    \int d\chi d\chi^* g(z) h(z^*) \psi_{gv_{\parallel}}(r),
\end{equation}
and the angular power spectrum:
\begin{align}
    &C_{\ell}^{\rm B-Doppler}
    \propto
    \int_{-1}^1 d\mu \hspace{0.1 cm} P_{\ell}(\mu) \hspace{0.1 cm} C^{\rm B-Doppler}(\theta)  \\
    &= 
    \int_{-1}^1 d\mu \hspace{0.1 cm} P_{\ell}(\mu) 
    \int d\chi d\chi^* (1+z)^2 (1+z^*) b(z^*) \psi_{gv_{\parallel}}(r).
\end{align}
For a more detailed derivation, see \cite{Alvarez2016}.

\end{document}